# Mechanics of soft-body rolling motion without external torque

Xudong Liang[1,*], Yimiao Ding[2], Zihao Yuan[6], Junqi Jiang[1], Zongling Xie[7,8], Peng Fei[9], Yixuan Sun[10], Guoying Gu[5,6], Zheng Zhong[1], Feifei Chen[5,6] †, Guangwei Si[7,8] †, Zhefeng Gong[2,3,4] ‡

[1] School of Science, Harbin Institute of Technology, Shenzhen 518055, China
[2] Department of Neurology of the Fourth Hospital and School of Brain Science and Brain Medicine, Zhejiang University School of Medicine, Hangzhou 310058, China
[3] Liangzhu Laboratory, MOE Frontier Science Center for Brain Science and Brain-machine Integration, State Key Laboratory of Brain-machine Intelligence, Zhejiang University, 1369 West Wenyi Road, Hangzhou 311121, China
[4] NHC and CAMS Key Laboratory of Medical Neurobiology, Zhejiang University, Hangzhou 310058, China
[5] State Key Laboratory of Mechanical System and Vibration, Shanghai Jiao Tong University, Shanghai 200240, China
[6] Robotics Institute, School of Mechanical Engineering, Shanghai Jiao Tong University, Shanghai 200240, China
[7] State Key Laboratory of Brain and Cognitive Science, Institute of Biophysics, Chinese Academy of Sciences, 15 Datun Road, Beijing 100101, China;
[8] University of Chinese Academy of Sciences, 19A Yuquan Road, Beijing 100049, China
[9] Huazhong University of Science and Technology, Optical and Electronic Information, Wuhan, 430074, China
[10] Zhejiang Lab, Hangzhou, 311121, China

The *Drosophila* larva, a soft-body animal, can bend its body and roll efficiently to escape danger. However, contrary to common belief, this rolling motion is not driven by the imbalance of gravity and ground reaction forces. Through functional imaging and ablation experiments, we demonstrate that the sequential actuation of axial muscles within an appropriate range of angles is critical for generating rolling. We model the interplay between muscle contraction, hydrostatic skeleton deformation, and body-environment interactions, and systematically explain how sequential muscle actuation generates the rolling motion. Additionally, we constructed a pneumatic soft robot to mimic the larval rolling strategy, successfully validating our model. This mechanics model of soft-body rolling motion not only advances the study of related neural circuits, but also holds potential for applications in soft robotics.

The development of wheels with rolling motion is the hallmark of civilization. However, rolling motion is not easy for most animals, which are only employed under emergent situations [1]. For example, wheel spiders [2], and web-toed salamanders [3] can curl their body into a hoop or sphere shape, and roll down a slope or surface driven by gravity or winds (Fig. 1(a)). Nevertheless, the rolling is passive without control of their locomotion. Animals such as caterpillars [4] have developed active rolling by bending their bodies and releasing the energy to generate an abrupt forward rolling; see Fig. 1(b). The single push drives the active rolling by the torque induced by the imbalance between the ground interaction and gravity. However, a continuous rolling motion is not sustained, given the lack of sustainable supplies of the external torque.

*Drosophila* larvae adopt a lateral rolling strategy to escape upon nociceptive stimuli [5]. In contrast to the head-to-tail type of rolling, the larvae generate a "C-shape" bending of its body ("C-bend") and rotate around the center of its cross-sections continuously with a varying rolling speed (Movie S1, Fig. S3 in Supplementary Materials [6]). The curvature of the bent larva body during rolling can be a constant or moderately oscillate (Fig. 1(c), Movie S2), indicating that the rolling motion is not powered by the release of elastic energy stored in bending deformation. In addition, the larvae can roll both towards or against the opening direction of the "C-bend", different from the rolling of the thermally activated polymeric bar [7-12], or rolling micro swarms enabled by magnetic fields [13,14], which indicates the driving toque of larval rolling is independent of the body's bending deformation direction.

Traditionally, larval rolling is proposed to be realized through the imbalance between ground reaction forces and gravity, like in most mammals and robots [15,17-19]. Yet the tangential force generated during rolling is about 1mN (Table S1 in Supplementary Materials [6]), much larger than the gravity at ~0.01mN [20] and ground reaction force at 0.01~0.07mN [21]. Gravity cannot be the dominant driving force for the larvae to roll. Indeed, rolling motion is observed in the larvae placed on a platform flipped upside down (Movies S3 and S4). This discrepancy prompts us to disclose the origin of the driving torque for larval rolling from biological experiments.

We monitor the muscle morphologies and activities in the rolling motions of transgenic fruit fly larvae, labeling all muscle cells with the activity indicator protein [1,22-24]. The larvae are placed in a microfluidic chamber in a water bath under a light-sheet microscope (Supplementary Materials [6], Methods). When water is heated to 40 ºC, rolling is induced, and the activities and morphology of the muscles are tracked (Fig. 2a, Movie S5). The muscle system in each of the 11 segments of the larvae is primarily composed of axial muscles around the body and circular muscles on the lateral sides (see Fig. 2(b)). We tracked the position, length, and calcium signal of each axial and circular muscle. As shown in Fig. 2(b), when the muscles, axial or circular, move toward the inner side of the "C-bend", their activity levels increase and peak on the inner side of the "C-bend". Therefore, body wall muscles are sequentially activated when the larva rolls. The progression of muscle activation is opposite to the rolling direction, which is independent of the opening direction of the "C-bend". These observations agree with the recent studies [15,16]. Meanwhile, the length of axial muscles clearly shows an increasing trend from the inner side to the outer side of the "C-bend". In contrast, the length of the circular muscles remains largely constant throughout the rolling process, as shown in Fig. 2(b), suggesting that axial muscles and circular muscles contribute differently to the rolling behavior.

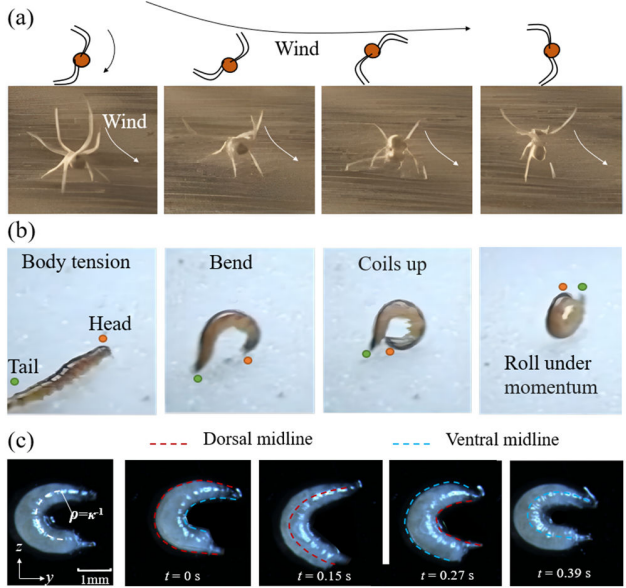

FIG. 1. Rolling motions in nature. (a) Wheel spiders with passive rolling driven by wind [2]. (b) Active rolling in caterpillar by releasing elastic energy [4]. (c) Continuous lateral rolling of *Drosophila* larva with the bent body. Scale bar: 1 mm.

To determine the muscles required for continuous rolling, we laser-ablated a group of axial and circular muscles and monitored the behavioral outcome (Movie S6 and Supplementary Materials [6]). As shown in Fig 2(c), when circular muscles on one side in all segments are ablated, the larvae still can complete the rolling cycles like the intact larvae, but with slower velocity. On the other hand, the larvae with dorsal or ventral axial muscles in all segments ablated are unable to complete a full rolling cycle. Notably, the larvae stop rolling as the ablated axial muscles arrive at about $\pi/2$ from the inner side of the "C-bend", where these muscles should have been activated in a continuous rolling cycle. Thus, the sequential activation of the axial muscles, instead of the circular ones, is necessary for larval rolling. In the following analysis, we focus only on the axial muscles.

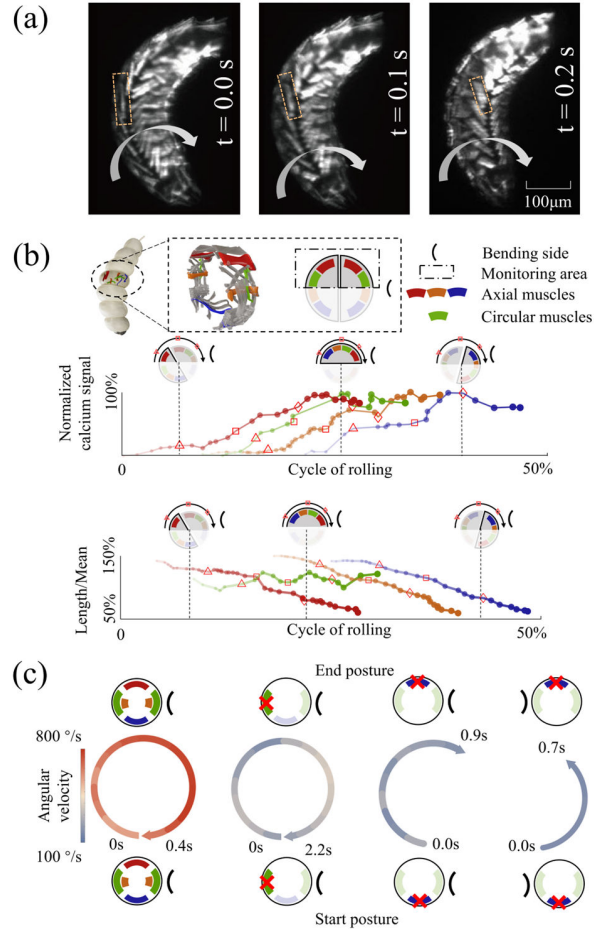

FIG. 2. Sequential contraction of axial muscles drives *Drosophila* larval rolling. (a) Measurement of muscle activities during inward rolling of a representative larva. (b) Changes in muscle activity level and length in inward rolling. Schematics for the larval body wall muscles are shown at the top. (c) Rolling of larvae with axial or circular muscle ablations.

Imaging and ablation experiments reveal the relationship between muscle actuation and rolling progression. When the axial muscles on one lateral side contract, the larval body bends into a C-shape. Then, the muscles relax, and the body tends to recover from the bending. Meanwhile, the adjacent axial muscles contract and bend the body toward the adjacent direction, so that change the opening of the "C-bend". Thus, the sequential muscle contraction and relaxation continuously change the body's bending direction, driving the body to roll, as shown in Fig. 3a.

To obtain the driving torque caused by sequential actuations of axial muscles, we derived a model to analyze the process of the deformation of the larval body (Supplementary Materials [6]). As shown in Fig 3(a), the larval body is simplified as a cylindrical tube with uniform circular cross-sections. The membrane is modeled as linear elastic material with Young's modulus $E_0$. The larvae maintain a hydraulic skeleton structure with internal pressure $p$, with an axial strain $\varepsilon_0=pR/2E_0h$, where $R$ and $h$ are the radius and thickness of the larvae cross-section. The axial muscles are evenly distributed around the body. The location of the muscle relative to the body is defined by the angle $\theta$, with $\theta=0$ lying at the inner side of the "C-bend". When the axial muscle at $\theta=0$ actives, it undergoes an isotonic contraction with a strain of $\Delta\varepsilon$, leading to the axial strain $\varepsilon_1$. With internal pressures, the contraction leads to bending deformation of the larvae body. The strain at $\theta=0$ is $\varepsilon_1=\varepsilon_0-\Delta\varepsilon$, and it becomes $\varepsilon_0$ at $\theta=\pi/2$, and $\varepsilon_2=\varepsilon_0+\Delta\varepsilon$ at $\theta=\pi$ (Fig. 3(a)). Assuming the strain changes with the angle $\theta$ linearly, the moment generated by the stresses in the deformed cross-sections is $M=8E_0R^2h\Delta\varepsilon/\pi$ (Supplementary Materials [6]), which is balanced by the moment generated by the force in the active axial muscle: $F_m=8E_0Rh\Delta\varepsilon/\pi$.

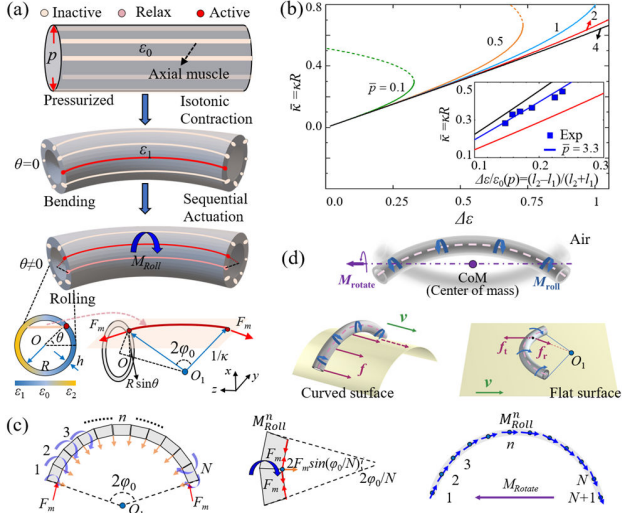

FIG. 3. Mechanical model for the rolling motion. (a) The axial muscle activities and the resulting bending and rolling. (b) The relation between the axial muscle contraction strain and the body curvature. Each mark represents an individual trial. (c) The rolling moment and the rotate moment induced by the axial muscle in the segmental body. (d) The larvae rolling motion when floated in the air without external forces (top), on a curved surface (bottom left), and on a flat surface (bottom right).

When the axial muscle at $\theta=0$ becomes inactive, the body tends to recover from bending. Meanwhile, the adjacent axial muscles at $\theta\neq0$ need to be activated to drive rolling. The muscle, with contraction force $F_m$, will induce a moment at the center of mass as $\sum \vec{r}\times\vec{F}_m$, where $\vec{r}$ is the vector between the center of curvature and the axial muscle force (Fig. 3(a) and Supplementary Materials [6]),

$$\vec{M}_{Roll}=-2F_mR\sin\varphi_0\sin\theta\vec{e}_z \quad (1)$$

where $2\varphi_0$ measures the arc of the bent body. The rolling direction is determined by the sign of $\theta$, which represents the progression of the axial muscle activation, but is independent of the opening direction of the "C-bend". If the axial muscle in the upper part with $0<\theta<\pi/2$ is activated, the larvae roll toward the opening direction of the "C-bend". Otherwise, the activated axial muscle in the lower part with $-\pi/2<\theta<0$ drives the larvae to roll in the opposite direction. If the axial muscle located at $|\theta|>\pi/2$ is activated, the larvae will reverse the "C-bend" without rolling.

Body curvature is an important factor in determining the rolling torque. Based on the model of the thin-walled cylindrical tube with internal pressures [25], we can obtain the relationship between curvature and the moment applied to the cross-section, $M=\frac{\pi E_0 R^2 h}{1-\nu^2}\left[\bar{\kappa}-\frac{3\bar{\kappa}^3}{8\bar{p}}\right]$, where $\bar{\kappa}=\kappa R$ and $\bar{p}=\frac{pR(1-\nu^2)}{E_0h}=2(1-\nu^2)\varepsilon_0$ are the scaled curvature and internal pressure, respectively, $\nu$ is the Poisson's ratio of the body (see derivations in Supplementary Materials [6]). Combining it with equation (1), the axial muscle contraction strain $\Delta\varepsilon$ and the curvature of the larvae is related as,

$$\Delta\varepsilon=\frac{\pi^2}{8(1-\nu^2)}\left[\bar{\kappa}-\frac{3\bar{\kappa}^3}{8\bar{p}}\right] \quad (2)$$

Based on Eq. (2), we plot the curvature $\bar{\kappa}$ and contraction strains $\Delta\varepsilon$ with different internal pressure $\bar{p}$. See Fig. 3(b). The larval body curvature in rolling increases with $\Delta\varepsilon$ and $\bar{p}$ (solid lines), and the dashed lines represent the unstable bending with kinking deformation [25]. Therefore, larger axial muscle contraction can boost the body's curvature. Higher internal pressure, induced by the hydrostatic skeleton structure, can make the larval body afford a higher curvature limit. Based on our model, we can estimate the internal pressure with the curvature and muscle contraction strain measured from experiments. The strain induced by axial muscle contraction can be measured as: $\Delta\varepsilon=\frac{\bar{p}(l_2-l_1)}{2(1-\nu^2)(l_2+l_1)}$, where $l_1$ and $l_2$ are the ventral muscle lengths at $\theta=0$ and $\theta=\pi$, respectively [26]. Considering Poisson's ratio $\nu=0.5$, we found the dimensionless internal pressure $\bar{p}=3.3$, which is shown in the inset of Fig. 3(b). Given the larvae skin thickness ratio $h/R=0.1$ (thickness 30 μm and radius 300 μm) and the elastic moduli of the skin $E_0$ ranging between 1.8 to 3.2 MPa (Supplementary Materials [6]), we estimate the internal pressure of the rolling larvae ranging from 0.8 to 1.4 MPa.

We further extend the model to consider the segmental larval body. For simplification, the body is evenly divided into $N$ segments. In each segment, uniformly distributed axial muscles are attached to the body wall around the segmental junctions [27]. Consider the plane of $x=R\sin\theta$, at each segment junction, the axial muscles form a resultant force pointing to the center of the "C-bend", and produce a moment through the center of the local cross-section $M_{Roll}^n=2F_mR\sin\frac{\varphi_0}{N}\sin\theta$ (see Fig3(c)). This distributed moment field drives the larval body roll through a curved axis, with a total torque $M_{Roll}=NM_{Roll}^n$. The axial muscles at the tip and end also generate a torque to rotate through the center of mass, changing the opening direction of the "C-bend" (Fig. 3(d), top): $M_{Rotate} = 2F_m R \cos\frac{\varphi_0}{N} \sin\varphi_0 \sin\theta$. Suppose a larva floats in the air, without external forces. In that case, sequential axial muscle contraction can drive the body to roll along the curved midline and rotate as a whole, changing the opening direction of the "C-bend". The rotation direction is opposite to rolling. Overall, the angular momentum is conserved (Fig. 3(d) and Supplementary Materials [6]).

When the larva is placed on a surface, the interaction between the body and the surface drives the translational movement. If the surface is a curved tube, and the curvature matches the shape of the "C-bend", the larva can move along the tube while rolling [28]. To initiate rolling, $M_{Roll}$ needs to surpass a threshold $M_{th}$, the resistance moment caused by the contact surface. Under non-slip conditions that the translational acceleration matches with the angular acceleration of rolling, we can derive the friction $f = \frac{2(M_{Roll}-M_{th})}{3R}$, pointing to the direction of translational movement (Fig 3(d)). When the segmental larvae roll on a flat surface, at each point of contact, the movements can be decomposed into rotational and translational movements. Accordingly, the friction force can be divided into: $f_r$ opposes the rolling and $f_t$ opposes the translational motion. Only at the middle point, the rolling can be non-slipping, where we can derive the relationship between $f_r$ and $f_t$: $f_r\left(\frac{\sin\varphi_0}{2\varphi_0} + 1\right) - \frac{f_t}{2}\left(1 + \frac{2\sin\varphi_0}{\varphi_0}\right) = \frac{M_{Roll}-M_{th}}{R}$. If the larva moves with a constant speed, we can further get: $f_r = f_t \frac{\varphi_0}{\sin\varphi_0} = 3f/2(1 - \frac{\sin^2\varphi_0}{\varphi_0^2})$. The driving force of translational motion decays with $\varphi_0$, or the curvature of bending. Only at the middle point of the arc does the rolling friction fully contribute to the translational movement. For segments away from this point, the rolling friction contributes less and less to translational motion.

Inspired by the proposed rolling mechanism of *Drosophila* larvae, we developed a pneumatic soft-rolling robot, as shown in Fig. 4(a). The robot is constrained by fibers in the hoop direction, and actuated by four evenly distributed chambers inside the body (Supplementary Materials [6]). When only one chamber is pressurized, it elongates and makes the robot bend toward the opposite side. The relationship between the bending angle and the applied pressure is shown in Fig. 4(b). As we pressurize the adjacent chamber and deflate the pressurized one at the same time, the robot will rotate toward the outer side of the "C-bend", leading to directional rolling locomotion. The soft robot can roll continuously when the four chambers are pressurized in sequence, as shown in Fig. 4(c) and Movie S7. We note that the soft robot's chambers are the counterpart of *Drosophila* larvae's muscles. However, since the chambers operate by elongation instead of contraction in muscles, the robot's rolling direction is opposite to that of the larvae.

In particular, the rolling motion of the pneumatic soft robot is highly robust even in the presence of damaged actuators. Based on our rolling model, converting the curvature of the "C-bend" and changing the positions of activated muscles between the upper and lower part can change the sign of $\sin\varphi_0$ and $\sin\theta$ simultaneously, without altering the rolling direction. The proof-of-concept experiment is first demonstrated in our rolling robot. When one of the pneumatic actuators is deactivated, the robot stops as the damaged actuator rotates to $\theta = \pm\pi/2$. By reversing the curvature and sequentially actuating other chambers, the robot continues to roll in the same direction (Fig. 4(d) and Movie S8). Interestingly, larvae with the axial muscle ablated can also keep the rolling direction by changing the opening direction of the "C-bend", although it is rarely observed, at a rate of no more than 1 in 20 (Movie S6).

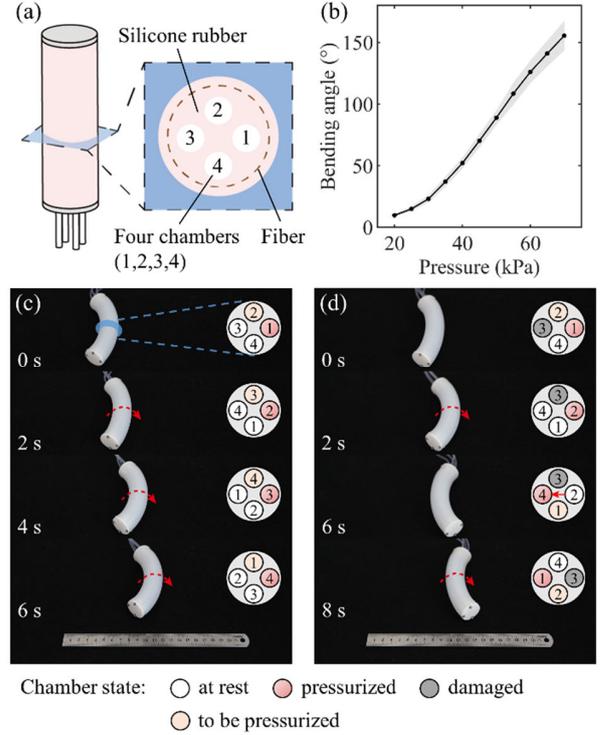

FIG. 4. Pneumatic soft rolling robot. (a) Design of the soft robot. (b) The relationship between the bending angle and pressure. (c) Demonstration of continuous rolling achieved by sequential actuation. (d) With one chamber damaged, the rolling continues by switching the bending direction (from 2 s to 6 s).

Without external driving torque, how can the larvae roll by itself? This difficulty arises from the incorrect application of experience with near-rigid objects to deformable active materials. Just like the falling cat problem, a cat can manipulate its rotational inertia and flip the whole body without external torque [29, 30]. Larvae can freely deform with no skeleton constraint. Their rolling behavior is ideal for studying the dynamical properties of deformable active materials. In this letter, we showed that the sequential activation of axial muscles is sufficient to generate a rolling moment independent of gravity and reaction force from the environment. Based on our model, the rolling direction is solely determined by the progression direction of muscle actuation. We further validated the proposed mechanism by developing a rolling soft robot. Our work opens an avenue for understanding the mechanics of soft-bodied animal locomotory behaviors, developing bio-inspired robots, and shedding light on new control strategies for locomotion.

This work is supported by the Major program of the National Natural Science Foundation of China (T2293720/T2293722, T2293721, T2293725). X. Liang acknowledges the support of the National Natural Science Foundation of China (12322207) and the Fundamental Research Funds for the Central Universities (HIT. OCEF. 2022037).

X. L., D. Y., and Z. Y. contributed equally to this work. X. L., F. C., G. S., and Z. G. conceived the idea and supervised the project.

* liangxudong@hit.edu.cn.
† ffchen@sjtu.edu.cn
† si@ibp.ac.cn
‡ zfgong@zju.edu.cn

[24] D. Li, F. Li, P. Guttipatti, and Y. Song, A Drosophila in vivo injury model for studying neuroregeneration in the peripheral and central nervous system, J Vis Exp (2018).

[25] L. Qiu, J. W. Hutchinson, and A. Amir, Bending instability of rod-shaped bacteria, Phys Rev Lett **128**, 058101 (2022).

[26] The strains in ventral axial muscle change from $\varepsilon_2=\varepsilon_0+\Delta\varepsilon$ to $\varepsilon_1=\varepsilon_0-\Delta\varepsilon$ as it rotates from $\theta=\pi$ to 0. The muscle lengths are $l_2=\varepsilon_2 l_0=(\varepsilon_0+\Delta\varepsilon)l_0$ and $l_1=\varepsilon_1 l_0=(\varepsilon_0-\Delta\varepsilon)l_0$, where $l_0$ is the initial length of the muscles. Therefore, the contraction strain of the axial muscle is $\Delta\varepsilon = (l_2 - l_1)\varepsilon_0/(l_2 + l_1) = \bar{p}(l_2 - l_1)/2(1 - v^2)(l_2 + l_1)$.

[27] J. Krzemien, CC. Fabre, J. Casal, and P. A. Lawrence, The muscle pattern of the Drosophila abdomen depends on a subdivision of the anterior compartment of each segment. Development **139**, 75 (2012)

[28] Z. Deng, K. Li, A. Priimagi, and H. Zeng, Light-steerable locomotion using zero-elastic-energy modes. Nat. Mater. **15**, 1688 (2024)

[29] T. Kane and M. Scher, A dynamical explanation of the falling cat phenomenon. International Journal of solids and structures **5**, 663 (1969).

[30] R. Montgomery,. Gauge theory of the falling cat. Fields Inst. Commun **1**, 1090 (1993).
6/6